# Computer modelling of hafnium doping in lithium niobate

R. M. Araujo (1), M. E. G. Valerio (2), R. A. Jackson (3)*

(1) Chemistry Department, Pio Decimo College, Campus III, Aracaju-SE, Brazil
(2) Physics Department, Federal University of Sergipe, Campus Universitário, 491000-000 São Cristovão-SE, Brazil
(3) Lennard-Jones Laboratories, School of Physical and Geographical Sciences, Keele University, Keele, Staffordshire ST5 5BG, UK

* Corresponding author, e-mail: r.a.jackson@keele.ac.uk

**Abstract**

Lithium niobate, $LiNbO_3$, is an important technological material with good electro-optic, acousto-optic, elasto-optic, piezoelectric and nonlinear properties. Doping $LiNbO_3$ with hafnium, Hf has been shown to improve the resistance of the material to optical damage. Computer modelling provides a useful means of determining the properties of doped and undoped $LiNbO_3$, including its defect chemistry, and the effect of doping on the structure. In this paper, Hf doped $LiNbO_3$ has been modelled, and the final defect configurations are found to be consistent with experimental results.

## 1. Introduction

Lithium niobate, $LiNbO_3$, is a material with many important technological applications that result from its diverse physical properties [1-4]. Laser induced optical damage or so-called photorefraction was first observed in $LiNbO_3$ and $LiTiO_3$ crystals at the Bell Laboratories [5]. This effect can be utilized for holographic information storage and optical amplification; however, it hinders the usage of $LiNbO_3$ in frequency doublers, Q-switchers and optical waveguides, so ways of minimising this optical damage have been sought actively. Kokanyan et al [6] reported that the light induced birefringence changes of $LiNbO_3$ crystals doped with 4 mol% of $HfO_2$ were comparable to that of 6 mol% MgO doped crystals, indicating that Hf doping is effective in resisting optical damage.

Much useful information about lithium niobate and its defect properties can be obtained by computer modelling, based on the description of interactions between ions by effective potentials. Previous papers have reported the derivation of an interatomic potential for $LiNbO_3$ [7], the doping of the structure by rare earth ions [8, 9], doping with Sc, Cr, Fe and In [10], and metal co-doping [11]. These papers show that modelling can predict the energetically optimal locations of the dopant ions, and calculate the energy involved in the doping process, making it a suitable method to study Hf-doped lithium niobate, with the aim of establishing the optimal doping site and charge compensation scheme.

## 2. Methodology

In this paper, use is made of the lattice energy minimisation method, in which the lattice energy of a given structure is calculated, and the structure varied until a minimum in the energy is found. This approach has been applied to a wide range of inorganic materials, with specific applications to $LiNbO_3$ reported in references [7]-[11]. The method makes use of interatomic potentials to describe the interactions between ions in the solid, as described in the next section, 2.1. Defects in solids are modelled using the Mott-Littleton method [12] which is described in section 2.2. All calculations were performed using the GULP code [13].

## 2.1 Interatomic potentials

In this paper use has been made of a previously derived potential for $LiNbO_3$ [7], and a potential fitted to the structure of $HfO_2$. In both cases, a Buckingham potential is employed, supplemented by an electrostatic interaction term:

$$V = \frac{q_1 q_2}{r} + A\exp\left(\frac{-r}{\rho}\right) - Cr^{-6}$$

In this potential, $q_1$ and $q_2$ are the charges on the interacting ions separated by a distance r, and A, ρ and C are parameters that are fitted empirically.

The derivation of potentials for $LiNbO_3$ and $HfO_2$ are considered separately below.

### 2.1.1 $LiNbO_3$

Full details of the derivation of the $LiNbO_3$ potential are given in reference [7], but they will be summarised here. The potential was derived empirically by simultaneously fitting to the structures of $LiNbO_3$, $Li_2O$ and $Nb_2O_5$. The $O^{2-}$ - $O^{2-}$ potential obtained by Catlow [14] was retained as this is widely used in many other oxides. The $O^{2-}$ ion was described using the shell model [15], and a 3 body potential was used to model the interactions between niobium ions and nearest oxygen neighbours, which takes the form:

$$V_{3\ body} = ½\ k\ (\theta - \theta_0)^2$$

In this equation $\theta_0$ is the equilibrium bond angle and $k_\theta$ is the bond-bending force constant.

The potential parameters are given in table 1 below:

Table 1: Potential parameters for $LiNbO_3$ [7]

| **Interaction** | **A (eV)** | **ρ (Å)** | **C (eV Å$^6$)** |
|---|---|---|---|
| $Nb_{core}$–$O_{shell}$ | 1425.0 | 0.3650 | 0.0 |
| $Li_{core}$–$O_{shell}$ | 950.0 | 0.2610 | 0.0 |
| $O_{shell}$–$O_{shell}$ | 22764.0 | 0.1490 | 27.88 |
| **Shell parameters** | **Shell charge, Y (\|e\|)** | | **Spring constant, $k_r$ (eV Å$^{-2}$)** |
| $O^{2-}$ | -2.9 | | 70.0 |
| **3 body parameters** | **Force constant, $k_\theta$ (eV rad−2)** | | **Equilibrium angle, $\theta_0$** |
| $O_{shell}$–$Nb_{core}$–$O_{shell}$ | 0.5776 | | 90.0 |

A comparison of experimental [16] and calculated lattice parameters of $LiNbO_3$ can be found in table 2, showing that the derived potential reproduces the structural parameters to within a few percent.

Table 2: Comparison of experimental [16] and calculated lattice parameters for $LiNbO_3$

| **Parameter** | **Experimental** | **Calculated (0 K)** | **Δ%** | **Calculated (295 K)** | **Δ%** |
|---|---|---|---|---|---|
| a=b (Å) | 5.1474 | 5.1559 | 0.17 | 5.1868 | 0.77 |
| c (Å) | 13.8561 | 13.6834 | 1.24 | 13.7103 | 1.05 |

2.1.2 $HfO_2$

A potential was derived for $HfO_2$ by fitting to its structure [17]. The potential parameters are given in table 3 (with the $O^{2-}$ shell parameters having the same values as in $LiNbO_3$), and the agreement between calculated and experimental lattice parameters calculated at 0K and 293K is shown in table 4. As is seen from the Δ% values, good agreement is obtained using this potential.

Table 3: Interionic potentials obtained from a fit to the $HfO_2$ structure [17]

| **Interaction** | **A (eV)** | **ρ (Å)** | **C (eV Å$^6$)** |
|---|---|---|---|
| $Hf_{core}$-$O_{shell}$ | 1413.54 | 0.3509 | 0.0 |
| $O_{shell}$-$O_{shell}$ | 22764.0 | 0.1490 | 27.88 |

Table 4: Comparison of calculated and experimental lattice parameters

| **Parameter** | **Experimental [17]** | **Calculated (0K)** | **Δ%** | **Calculated (295K)** | **Δ%** |
|---|---|---|---|---|---|
| a=b=c (Å) | 5.084000 | 5.084236 | 0.00 | 5.087119 | 0.06 |

2.2 Defect calculations

The calculations are carried out using the Mott–Littleton method [12], in which point defects are considered to be at the centre of a region in which all interactions are treated explicitly, while approximate methods are employed for regions of the lattice more distant from the defect. In practice, this involves placing the $Hf^{4+}$ ion at either the $Li^+$ or $Nb^{5+}$ site, along with a range of charge compensating defects, as listed below, using schemes (i) and (ii) suggested by Li et al [19], plus a further 5 schemes ((iii)-(vii)) proposed here:

(i) An $Hf^{4+}$ ion at a $Li^+$ site, with charge compensation by 3 $Li^+$ vacancies

(ii) An $Hf^{4+}$ ion at a $Li^+$ site, with charge compensation by 3 $Hf^{4+}$ ions at $Nb^{5+}$ sites

(iii) 4 $Hf^{4+}$ ions at $Nb^{5+}$ sites, with charge compensation by a $Nb^{5+}$ ion at a $Li^+$ site

(iv) An $Hf^{4+}$ ion at a $Nb^{5+}$ site, with charge compensation by a $Nb^{5+}$ ion at a $Li^+$ site and 3 $Li^+$ vacancies

(v) 2 $Hf^{4+}$ ions at $Nb^{5+}$ sites, with charge compensation by a $Nb^{5+}$ ion at a $Li^+$ site and 2 $Li^+$ vacancies

(vi) 3 $Hf^{4+}$ ions at $Nb^{5+}$ sites, with charge compensation by a $Nb^{5+}$ ion at a $Li^{+}$ site and 1 $Li^{+}$ vacancy

(vii) 2 $Hf^{4+}$ ions at $Nb^{5+}$ sites, with charge compensation by an $O^{2-}$ vacancy

**3. Results and Discussion**

The seven mechanisms described in section 2.2 have been written below as solid-state reactions, employing Kroger–Vink notation [20]:

$$HfO_2 + 4Li_{Li} \rightarrow Hf_{Li}^{\bullet\bullet\bullet} + 3V_{Li}^{'} + 2Li_2O \text{ (i)}$$

$$4HfO_2 + Li_{Li} + 3Nb_{Nb} \rightarrow Hf_{Li}^{\bullet\bullet\bullet} + 3Hf_{Nb}^{'} + \frac{1}{2}Li_2O + \frac{3}{2}Nb_2O_5 \text{ (ii)}$$

$$4HfO_2 + Li_{Li} + 4Nb_{Nb} \rightarrow 4Hf_{Nb}^{'} + Nb_{Li}^{\bullet\bullet\bullet\bullet} + \frac{1}{2}Li_2O + \frac{3}{2}Nb_2O_5 \text{ (iii)}$$

$$HfO_2 + 4Li_{Li} + Nb_{Nb} \rightarrow Hf_{Nb}^{'} + Nb_{Li}^{\bullet\bullet\bullet\bullet} + 3V_{Li}^{'} + 2Li_2O \text{ (iv)}$$

$$2HfO_2 + 3Li_{Li} + 2Nb_{Nb} \rightarrow 2Hf_{Nb}^{'} + Nb_{Li}^{\bullet\bullet\bullet\bullet} + 2V_{Li}^{'} + Li_2O + LiNbO_3 \text{ (v)}$$

$$3HfO_2 + 2Li_{Li} + 3Nb_{Nb} \rightarrow 3Hf_{Nb}^{'} + Nb_{Li}^{\bullet\bullet\bullet\bullet} + V_{Li}^{'} + 2LiNbO_3 \text{ (vi)}$$

$$2HfO_2 + 2Nb_{Nb} + O_O \rightarrow 2Hf_{Nb}^{'} + V_O^{\bullet\bullet} + Nb_2O_5 \text{ (vii)}$$

The energies corresponding to these reactions are defined as solution energies, $E_s$, and they are calculated as follows:

$$E_s = E(Hf_{Li}^{\bullet\bullet\bullet} + 3V_{Li}^{'}) + 2E_{latt}(Li_2O) - E_{latt}(HfO_2) \text{ (i)}$$

$$E_s = E(Hf_{Li}^{\bullet\bullet\bullet} + 3Hf_{Nb}^{'}) + \frac{3}{2}E_{latt}(Nb_2O_5) + \frac{1}{2}E_{latt}(Li_2O) - 4E_{latt}(HfO_2) \text{ (ii)}$$

$$E_s = E(4Hf_{Nb}^{'} + Nb_{Li}^{\bullet\bullet\bullet\bullet}) + \frac{3}{2}E_{latt}(Nb_2O_5) + \frac{1}{2}E_{latt}(Li_2O) - 4E_{latt}(HfO_2) \text{ (iii)}$$

$$E_s = E(Hf_{Nb}^{'} + Nb_{Li}^{\bullet\bullet\bullet\bullet} + 3V_{Li}^{'}) + 2E_{latt}(Li_2O) - E_{latt}(HfO_2) \text{ (iv)}$$

$$E_s = E(2Hf_{Nb}^{'} + Nb_{Li}^{\bullet\bullet\bullet\bullet} + 2V_{Li}^{'}) + E_{latt}(Li_2O) + \mathrm{E}_{latt}(LiNbO_3) - 2E_{latt}(HfO_2) \text{ (v)}$$

$$E_s = E(3Hf_{Nb}^{'} + Nb_{Li}^{\bullet\bullet\bullet\bullet} + V_{Li}^{'}) + 2E_{latt}(LiNbO_3) - 3E_{latt}(HfO_2) \text{ (vi)}$$

$$E_s = E(2Hf_{Nb}^{'} + V_O^{\bullet\bullet}) + E_{latt}(Nb_2O_5) - 2E_{latt}(HfO_2) \text{ (v)}$$

Lattice energies, $E_{latt}$, required to calculate the solution energies, are given in table 5. Table 6 gives the formation energies of the bound defects (the first term in the above equations). Table 7 gives the solution energy for each scheme (determined using the expressions above), and it is noted that the lowest energy corresponds to scheme (ii), where the $Hf^{4+}$ ion is found at both cation sites (self-compensation). This is confirmed by experimental data [18, 19].

Table 5: Lattice energies used in the solution energy calculations (eV).

| Structures | 0 K | 293 K |
|---|---|---|
| **$LiNbO_3$** | -174.57 | -174.66 |
| **$Li_2O$** | -33.16 | -32.92 |
| **$Nb_2O_5$** | -314.37 | -313.99 |
| **$HfO_2$** | -110.39 | -110.45 |

Table 6: Defect formation energies, in eV, for the bound defect

| Defect | Scheme (i) | | Scheme (ii) | | Scheme (iii) | | Scheme (iv) | | Scheme (v) | | Scheme (vi) | | Scheme (vii) | |
|---|---|---|---|---|---|---|---|---|---|---|---|---|---|---|
| T(K) | 0K | 293K | 0K | 293K | 0K | 293K | 0K | 293K | 0K | 293K | 0K | 293K | 0K | 293K |
| $Hf^{4+}$ | -36.03 | -36.35 | 52.51 | 52.27 | 53.34 | 53.11 | -33.85 | -34.01 | -2.61 | -2.73 | 25.40 | 25.07 | 97.82 | 97.64 |

Table 7: Solution energies, in eV, for the bound defect (per dopant ion)

| Defect | Scheme (i) | | Scheme (ii) | | Scheme (iii) | | Scheme (iv) | | Scheme (v) | | Scheme (vi) | | Scheme (vii) | |
|---|---|---|---|---|---|---|---|---|---|---|---|---|---|---|
| **T(K)** | 0K | 293K | 0K | 293K | 0K | 293K | 0K | 293K | 0K | 293K | 0K | 293K | 0K | 293K |
| **$Hf^{4+}$** | 8.26 | 8.04 | 1.65 | 1.48 | 1.86 | 1.69 | 5.25 | 5.11 | 2.12 | 2.09 | 1.49 | 1.42 | 2.28 | 2.11 |

## 4. Conclusions

This paper has presented a computational study of $Hf^{4+}$ doped $LiNbO_3$. Solution energies have been calculated for seven possible mechanisms by which the $Hf^{4+}$ might be incorporated in the structure, and the lowest energy scheme, involving self-compensation, is shown to be consistent with experimental data.

**Acknowledgments**
The authors are grateful to CAPES and FINEP for financial support.